\documentclass[twocolumn, superscriptaddress,amsmath,amssymb,aps, prl,
showkeys,showpacs,nofootinbib] {revtex4-1}
\usepackage{graphicx}
\usepackage{color}
\usepackage{bm}
\usepackage{hyperref}
\usepackage{dcolumn}
\usepackage{slashed}

\begin{document}
\title{Remote Entanglement by Coherent Multiplication of Concurrent Quantum Signals}
\author{Ananda Roy}
\email{ananda.roy@yale.edu}
\affiliation{Department of Applied Physics, Yale University, PO BOX 208284, New Haven, CT 06511}
\author{Liang Jiang}
\affiliation{Department of Applied Physics, Yale University, PO BOX 208284, New Haven, CT 06511}
\author{A. Douglas Stone}
\affiliation{Department of Applied Physics, Yale University, PO BOX 208284, New Haven, CT 06511}
\author{Michel Devoret}
\affiliation{Department of Applied Physics, Yale University, PO BOX 208284, New Haven, CT 06511}

\begin{abstract}
Concurrent remote entanglement of distant, non-interacting quantum entities is a crucial function for quantum information processing. In contrast with the existing protocols which employ {\it addition} of signals to generate entanglement between two remote qubits, the continuous variable protocol we present is based on {\it multiplication} of signals. This protocol can be straightforwardly implemented by a novel Josephson junction mixing circuit. Our scheme would be able to generate provable entanglement even in presence of practical imperfections: finite quantum efficiency of detectors and undesired photon loss in current state-of-the-art devices.
\end{abstract}
\maketitle 

Generation of entangled states between spatially separated non-interacting quantum systems is an indispensable ingredient for large-scale quantum information processing \cite{Ekert_1991, Bennett_Wootters_1993, van_Enk_Zoller_1997, Briegel_Zoller_1998}. 
In particular,  {\it concurrent} remote entanglement, in which propagating quantum signals do not interact with both the systems under consideration, is a desirable feature of a scalable module-based architecture \cite{Duan_Monroe_2004, Jiang_Lukin_2007, Devoret_Schoelkopf_2013, Monroe_Kim_2014}.   

It is well known that a non-linear operation is necessary to achieve entanglement. Existing protocols for heralded concurrent remote entanglement employ linear optical elements in the processing stage (e.g. beam splitters), while the necessary nonlinearity is provided in the final stage by photon detection \cite{Cabrillo_Zoller_1999, KLM_2001, DLCZ, Chou_Kimble_2005, Moehring_Monroe_2007, Bernien_Hanson_2013, Hofmann_Weinfurter_2012}. While these methods relied on {\it addition} of signals using beam-splitters to erase `which qubit' information, our proposed method relies on {\it multiplication} of signals coming from each qubit to delete their local orientation. This multiplication is achieved by a new type of nonlinear signal processing. Josephson junction based superconducting circuit QED systems have access to strong, tunable, purely dispersive nonlinearities, making them natural candidates for implementing this protocol. In fact, sequential remote entanglement with linear microwave signal processing has already been performed using Josephson junction circuits \cite{Roch_Siddiqi_2014}. 
%



\begin{figure}
\centering
\includegraphics[width = 0.5\textwidth]{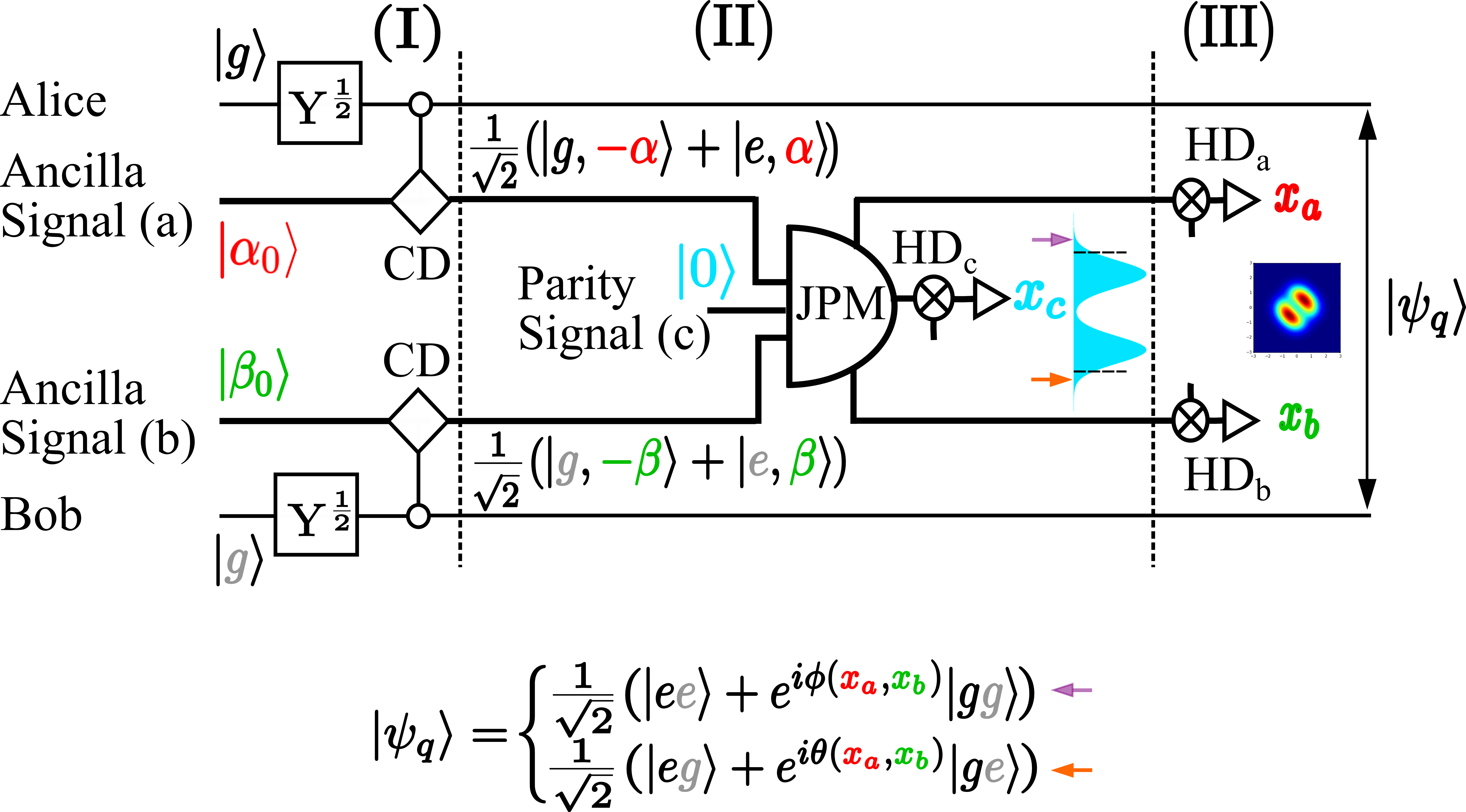}
\caption{\label{parity_entanglement} (color online) Remote entanglement protocol schematic. The first step of the protocol (I) consists of entangling two stationary qubits, Alice (in dark red) and Bob (in dark green), with two propagating modes (indicated by thicker lines), ancilla signal a and ancilla signal b  respectively, each initially prepared in a coherent state. This is achieved by first applying a $\pi/2$ rotation ($\rm{Y^{1/2}}$) on Alice (Bob), followed by a conditional displacement gate (CD) on Alice (Bob) and ancilla  signal a (ancilla signal b). In the next step (II), a nonlinear interaction of the ancilla signal a, ancilla signal b and parity signal c (implemented by the Josephson Parametric Multiplier (JPM), see below), followed by a homodyne detection of the parity signal mode, effectively realizes a joint two-qubit parity measurement. The resulting qubit-photon state has either even or odd joint qubit parity conditioned on the integrated homodyne current being in either lobe of the distribution (indicated by dashed lines). The last step (III), which disentangles the qubits from the photon states, comprises of homodyne measurements of the $\mathbf{a}$ and $\mathbf{b}$ modes, denoted by $\rm{HD}_a$ and $\rm{HD}_b$. Conditioned on the measurement outcome in (II) and (III), the  qubits are projected onto the even or odd Bell manifold, with a relative phase that depends on the measurement outcome in (III).}
\end{figure}

The first step of our protocol generates local entanglement \cite{Brune_Haroche_1996, Nemoto_Munro_2004} between a stationary superconducting qubit (for definiteness, a transmon qubit) and a propagating microwave mode (cf. (I) in Fig. \ref{parity_entanglement}) for each of Alice and Bob \cite{Hatridge_Devoret_2013}. 
Both Alice and Bob are initialized, using local $\pi/2$ rotations ($\rm{Y^{1/2}}$), to a superposition of their ground ($|g\rangle$) and excited ($|e\rangle$) states, given by: $(|g\rangle + |e\rangle)/\sqrt{2}$. Propagating modes, with coherent states of amplitude $\alpha_0$ and $\beta_0$, and temporal profile $e^{\kappa_a t/2}\cos(\omega_a t)\Theta(-t)$ and $e^{\kappa_b t/2}\cos(\omega_b t)\Theta(-t)$, are incident resonantly on two cavities, exciting their fundamental modes $\mathbf{A}$  and $\mathbf{B}$, with frequencies (decay rates) $\omega_a (\kappa_A)$ and $\omega_b (\kappa_B)$, with $\kappa_{a(b)}\ll\kappa_{A(B)}$. These modes interact dispersively through cross-Kerr interaction \cite{Blais_Schoelkopf_2004, Wallraff_Schoelkopf_2004} with Alice and Bob. This operation is referred to as the conditional displacement gate (CD). It imparts a qubit-state-dependent phase-shift on the outgoing microwave modes. The resultant entangled qubit-photon states output from Alice and Bob's cavities can be written as:  $(|e, \alpha\rangle + |g, -\alpha\rangle)/\sqrt{2}$ and $(|e, \beta\rangle + |g, -\beta\rangle)/\sqrt{2}$ \cite{Gambetta_Girvin_2008} with temporal profiles $ie^{\kappa_a t/2}\cos(\omega_a t)\Theta(-t)$ and $ie^{\kappa_b t/2}\cos(\omega_b t)\Theta(-t)$, respectively \cite{Gardiner_Collett_1985, Suppl}. Without loss of generality, we may assume $\alpha, \beta\in\Re$ and they need not be equal in our protocol. 

\begin{figure}
\includegraphics[width = 0.5\textwidth]{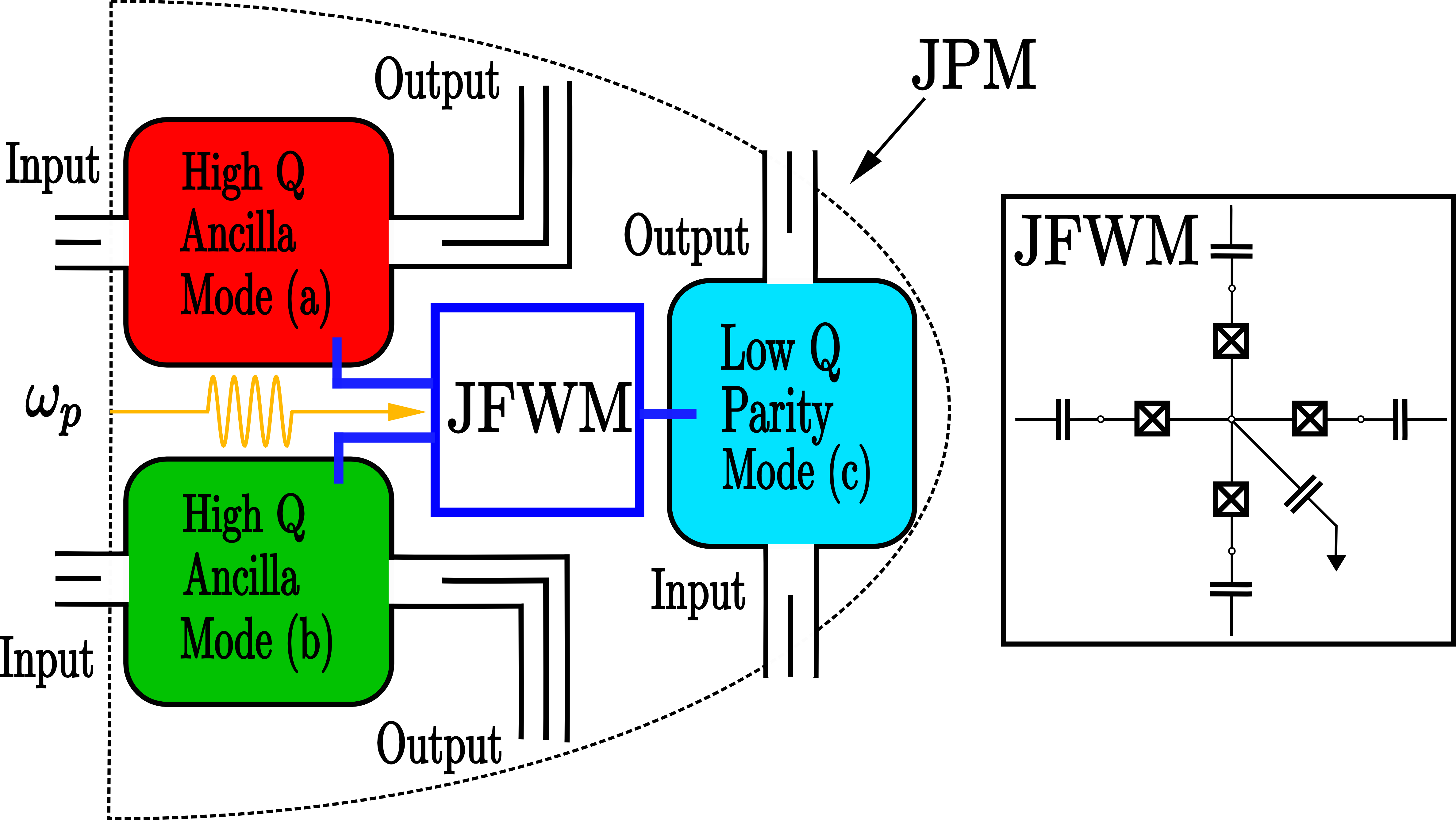}
\caption{ \label{schematic}  (color online) Schematic of the JPM of Fig. \ref{parity_entanglement}. The modes $\mathbf{a} (\omega_a), \mathbf{b} (\omega_b)$ and $\mathbf{c}(\omega_c)$ when pumped at $\omega_p= \omega_c-\omega_a -\omega_b$ (orange) participate in a non-linear, three-wave interaction $\mathbf{H}_{\rm{int}}/\hbar = ige^{-i\omega_p t}\mathbf{a}\mathbf{b}\mathbf{c}^\dagger + \rm{h.c.}$. This nonlinear mode mixing arises out of the Josephson Four Wave Mixer (JFWM). 
(Inset) The JFWM has four nominally identical Josephson junctions connected electrically as shown. The coupled system has four mutually orthogonal normal (electrical) modes \cite{Suppl}, which couple non-linearly, corresponding to the cavity modes $\mathbf{a,b,c}$ and the pump mode.}
\end{figure}

In the next step, we realize a joint two-qubit parity measurement by first capturing the propagating modes in resonators and then employing a nonlinear dissipation process. To that end, we introduce the Josephson Parametric Multiplier (JPM) (see step (II) in Fig. \ref{parity_entanglement}). The JPM comprises three resonators and a nonlinear four wave mixing element, the Josephson Four Wave Mixer (JFWM)  (Fig. \ref{schematic}). The three resonators have fundamental modes (frequencies, decay rates) $\mathbf{a} (\omega_a, \kappa_a), \mathbf{b} (\omega_b, \kappa_b)$ and $\mathbf{c} (\omega_c, \kappa_c)$. The outputs of cavity modes $\mathbf{A}$ and $\mathbf{B}$, after propagating through transmission lines, act as inputs to the $\mathbf{a}$ and $\mathbf{b}$ modes, respectively. Due to their particular temporal profiles, these flying modes are perfectly captured at $t=0$. A coupled two-mode dissipation is then turned on at $t=0$, which removes pairs of photons from the $\mathbf{a}$ and $\mathbf{b}$ modes at a rate $\kappa_{\rm{2ph}}$. This dissipation, mediated by the jump operator $\mathbf{a}\mathbf{b}$, is realized by the JFWM, together with the dissipation of the $\mathbf{c}$ mode in the following way. 

The JFWM consists of four nominally identical Josephson junctions, as shown in Fig. \ref{schematic} and has four interacting normal modes, which are negligibly shifted in frequency from the original modes $\mathbf{a,b,c}$, in the presence of a stiff, off-resonant pump mode with frequency chosen to be $\omega_p =  \omega_c-\omega_a - \omega_b$. Thus, under the rotating wave approximation, the mode-mixing arising out of the Josephson nonlinearity leads to an interaction Hamiltonian of the form $\mathbf{H}_{\rm{int}}/\hbar = ige^{-i\omega_pt}\mathbf{abc}^\dagger + \rm{h.c.}$, where $g$, the effective interaction strength, depends on the pump amplitude \cite{Suppl}. 
If the cavities are designed and pump strength is chosen such that
\begin{equation}		
\label{time_scale}
\kappa_a, \kappa_b\ll g, \kappa_{\rm{2ph}}\ll \kappa_c,
\end{equation}	
the JPM will provide unidirectional conversion: photons in modes $\mathbf{a}$ and $\mathbf{b}$ are converted into the $\mathbf{c}$ mode, which leaks out before it can be converted back into the $\mathbf{a}$ and $\mathbf{b}$, providing the desired two-photon dissipation channel, characterized by the decay rate $\kappa_{\rm{2ph}} = 4g^2/\kappa_c$ after adiabatic elimination \cite{Carmichael_2007}.
The nonlinear dissipation channel is monitored with a homodyne detection scheme, denoted by $\rm{HD}_c$, with phase angle  $\rm{arg}(g\alpha\beta)$, which measures the value of the integrated homodyne current $x_c$. By selecting outcomes $x_c$, which will fall on either lobe of the distribution, centered at $\pm2g\rm{min}(\alpha, \beta)^2/\kappa_c$, shown in (II) in Fig. \ref{parity_entanglement}, the qubit-photon state is projected on to the even or odd joint qubit-parity subspace. The error rate at this step can be controlled by selecting extremal outcomes, beyond some cut-off in each lobe of the distribution. At the end of the measurement, the two-photon dissipation is turned off by switching off the pump at $\omega_p$. 

The homodyne measurement of $\mathbf{c}$ in step (II), is followed in step (III) (Fig.~\ref{parity_entanglement}) by homodyne measurements of the modes  
$\mathbf{a}$ and $\mathbf{b}$, denoted by $\rm{HD}_a$ and $\rm{HD}_b$. This last pair of measurements is crucial  because, while the two-mode dissipation projects onto the even or odd qubit-parity subspace, the photons left over in the modes $\mathbf{a}$ and $\mathbf{b}$ after step (II) are in a two-mode squeezed state which remains entangled with the state of the qubits. Step (III) disentangles the qubits from these microwave modes, as follows.  

\begin{figure}[!h]
\includegraphics[width = 0.5\textwidth]{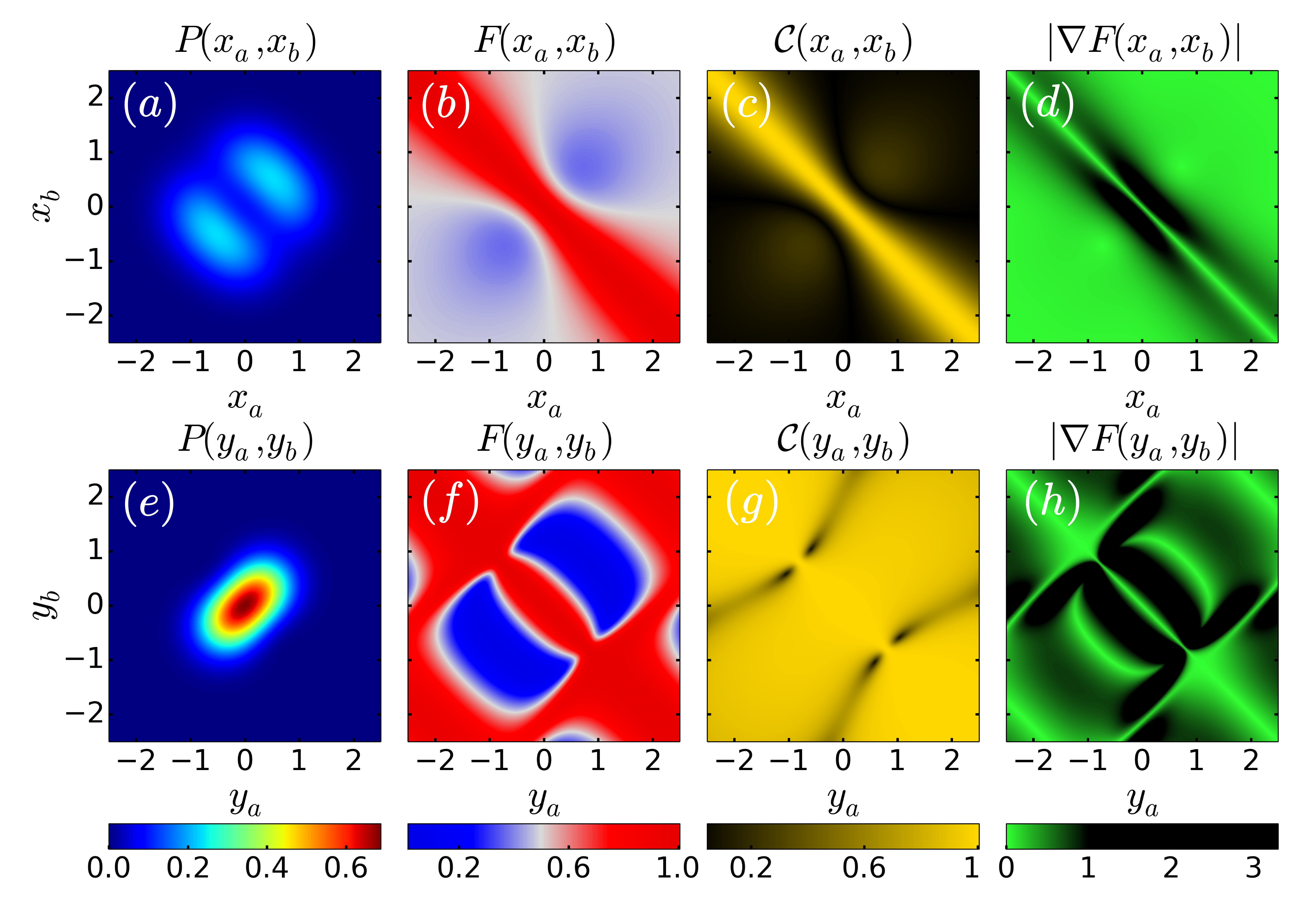}
\caption{ \label{fig_x_y_meas}  (color online) Probability distribution of outcomes, overlap (F) with Bell-state $|\Phi^+\rangle= (|ee\rangle + |gg\rangle)/\sqrt{2}$, concurrence (C) of the joint two-qubit system $\rho_q$ and gradient of overlap are plotted when the qubit-photon state is projected onto one with even two-qubit parity after the second step of the protocol. The top (bottom) row panels correspond to homodyne detection of the X (Y) quadratures of modes $\mathbf{a,b}$. We choose $\alpha= \beta = 0.75$ and assume perfect quantum efficiency and zero spurious photon loss. Panel (a) shows the probability of outcomes $P(x_a, x_b)$ for X measurements, corresponding to which, we see that the two-qubit state is projected on to $|\Phi^+\rangle$ for values around the diagonal $x_b = -x_a$ (panel (b)). For events occurring in the quadrant $x_a, x_b>0(<0)$, the two-qubit state is projected on to $|ee\rangle(|gg\rangle)$. Panel (c) shows the concurrence ${\cal C}(x_a, x_b)$ of $\rho_q$ for the different outcomes, which varies from $0$ (for $\rho_q$ being a separable state $|ee\rangle$ or $|gg\rangle$) to $1$ (in case of maximum entanglement). Panel (d) shows the gradient of the overlap as a function of $x_a, x_b$, which is zero (in green) for a narrow region around the line $x_a = -x_b$ where an entangled state is obtained. It changes rapidly on moving away from the line $x_a=-x_b$ and goes back to zero when the qubit state is projected on to $|ee\rangle$ or $|gg\rangle$.  Panel (e) shows the probability of outcomes $P(y_a, y_b)$ for Y measurements. For events around the line $y_a = -y_b$, the two-qubit state is once again projected on to $|\Phi^+\rangle$. However, for Y measurements, the phase of the generated Bell state varies continuously, depending on the particular outcome $(y_a, y_b)$, indicated by the existence of alternating bright and dark fringes in fidelity (panel (f)). For instance, along the line $y_a = y_b$, the two-qubit state oscillates continuously between $|\Phi^+\rangle$ and $|\Phi^-\rangle$. Panel (g) shows the concurrence ${\cal C}(y_a, y_b)$ of $\rho_q$, which is $\approx 1$ for all measurement outcomes, indicating generation of maximal entanglement for all outcomes $(y_a, y_b)$. The small regions of low entanglement are artifacts of the analytic approximation which replaces stochastic evolution of the system with a deterministic Lindblad evolution (see further discussion in the text). Panel (h) shows the gradient of overlap which is zero where the two-qubit state is projected on to $|\Phi^\pm\rangle$ and changes rapidly as the two-qubit state oscillates between $|\Phi^+\rangle$ and $|\Phi^-\rangle$. }
\end{figure}

Consider the case in which the two-qubit parity measurement projects the system to the even two-qubit parity subspace. 
The sign of the X quadrature measurements $x_a, x_b$ is correlated with the probability that the qubits are in the $|gg\rangle (x_a, x_b<0)$ or $|ee\rangle(x_a, x_b>0)$ states and only in certain regions of the $(x_a,x_b)$ plane along the line $x_a = -x_b$ are the two qubit states strongly entangled (Fig \ref{fig_x_y_meas}, upper panels).  Conversely, Y quadrature measurements will give results centered around $y_a=y_b =0$, are not correlated with the two qubit states, and do not distinguish between them; the result is that Y measurements always entangle the two qubits (with a relative phase which interpolates between the even and odd Bell states) (Fig. \ref{fig_x_y_meas}, lower panels). Similar reasoning holds for the odd two-qubit parity subspace outcomes of step (II). Hence, while it is possible to have reasonable success rate by making X measurements on modes $\mathbf{a}$ and $\mathbf{b}$, it is always preferable to measure the Y quadrature for optimal success rate.

While it is possible to perform complete stochastic master equation simulations of the protocol we have just outlined \cite{Suppl}, given the assumed separation of time scales (Eqn. \eqref{time_scale}), we have used in Fig. \ref{fig_x_y_meas} a simpler approximate, but accurate, model, which can be solved analytically and provides physical insight.  
Following the capture of the propagating microwave modes in signal resonators, the state of the system comprised of Alice, Bob and modes $\mathbf{a}, \mathbf{b}$, at $t=0$, is given by $\rho(t=0) = |\psi\rangle\langle\psi|$, where $|\psi\rangle = (|ee, \alpha, \beta\rangle + |gg, -\alpha, -\beta\rangle + |eg, \alpha, -\beta\rangle + |ge, -\alpha, \beta\rangle)/2$. Since the two-qubit parity measurement of step (II) depends only on unambiguously inferring which side of the distribution the outcome is on, for the purposes of analytic computation, we separately average over the different outcomes for the two lobes of the distribution (see Fig. \ref{parity_entanglement}). This amounts to replacing the stochastic evolution of the whole system by separate deterministic Lindblad evolutions for the even and odd qubit parity subspaces. During this evolution, the single photon losses of modes $\mathbf{a,b}$ are negligible due to Eqn. \eqref{time_scale}. The system density-matrix in even (odd) qubit-parity subspace $\rho_{e(o)}$ thus evolves according to: 
\begin{equation}
\label{lind_eqn}
\frac{d\rho_{e(o)}}{dt} = \kappa_{\rm{2ph}}{\cal D}(\mathbf{ab})\rho_{e(o)},
\end{equation}
where $\rho_{e(o)}(t=0) = |\psi_{e(o)}\rangle\langle\psi_{e(o)}|, |\psi_e\rangle=(|ee, \alpha, \beta\rangle + |gg, -\alpha, -\beta\rangle)/\sqrt{2}$, $|\psi_o\rangle=(|eg, \alpha, -\beta\rangle + |ge, -\alpha, \beta\rangle)/\sqrt{2}$ and ${\cal D}(O)\rho_{e(o)} = O\rho_{e(o)} O^\dagger - (O^\dagger O\rho_{e(o)} + \rho_{e(o)} O^\dagger O)/2$ is the Lindblad dissipation operator. 
The quasi-steady state at the end of this evolution, denoted by $\rho_{e(o)}^{\rm{qs}}$ \cite{Suppl}, subsequently evolves under the single photon loss of $\mathbf{a}$ and $\mathbf{b}$ that are monitored by $\rm{HD}_a$ and $\rm{HD}_b$. The resulting measurement of both $\Xi\in\{X,Y\}$ quadratures of $\mathbf{a}$ and $\mathbf{b}$ modes results in the system density matrix evolving to:
\begin{eqnarray}
\rho_{e(o)}^{\rm{qs}}\rightarrow \frac{{\cal M}_{\Xi}\rho_{e(o)}^{\rm{qs}}{\cal M}_{\Xi}^\dagger}{{\rm{Tr}}\big[{\cal M}_{\Xi}\rho_{e(o)}^{\rm{qs}}{\cal M}_{\Xi}^\dagger\big]},\ {\cal M}_{\Xi} = |\xi_a, \xi_b\rangle\langle \xi_a, \xi_b|.
\end{eqnarray}
The post-measurement two-qubit density matrix $\rho_q$ is computed by tracing out the modes $\mathbf{a}$ and $\mathbf{b}$. Conditioned on the outcomes in (II) and (III), the qubits are projected onto an entangled state in the subspace spanned by $\{|gg\rangle, |ee\rangle\}$ or $\{|eg\rangle, |ge\rangle\}$. The continuous nature of entanglement generation appears as a relative  complex amplitude of the two terms of the Bell state, which is determined by the measurement outcome in $\rm{HD_a}$ and $\rm{HD_b}$. 

In Fig. \ref{fig_x_y_meas}, we show the probability of outcomes, overlap with the Bell-state $|\Phi^+\rangle = (|ee\rangle + |gg\rangle)/\sqrt{2}$, concurrence  and gradient of the overlap for either X measurements (top row panels)  or Y measurements (bottom row panels) of the $\mathbf{a}$ and $\mathbf{b}$ modes. 
For X measurements, we see that the  majority of the events occur for either $x_a, x_b>0$ or $x_a, x_b<0$, hence projecting the qubit state onto product states $|ee\rangle$ or $|gg\rangle$. However, the (non-negligible) number of outcomes near the line $x_a = -x_b$, do project the qubit onto the entangle state $|\Phi^+\rangle$. Accordingly, the concurrence ${\cal C}(x_a,x_b)$ varies from $0$ (for $\rho_q$ separable) to $1$ (in case of maximum entanglement). The width of the region in phase-space where entanglement is generated is a function of $\alpha, \beta$ and decreases as $\alpha, \beta$ are increased. The rate of variation of entanglement is indicated by gradient of the overlap $|\Phi^+\rangle $ which varies most rapidly perpendicular to the line $x_a= -x_b$.
For Y measurements,in contrast,a maximally entangled state is generated for all outcomes, with the phase of the generated Bell state varying continuously in the form $|\Phi^\varphi\rangle=  (|ee\rangle +e^{i\varphi} |gg\rangle)/\sqrt{2}$, giving concurrence equal to unity at all points. An increase in $\alpha, \beta$ makes this variation more rapid. In this case the gradient of overlap is not a measure of entanglement, but just describes the variation of the phase, $\varphi (y_a,y_b)$.  A similar computation for the odd manifold shows similar results for X and Y measurements, with $|ee\rangle\rightarrow |eg\rangle, |gg\rangle\rightarrow|ge\rangle$ with the features in the Fig. \ref{fig_x_y_meas} rotated by $\pi/2$. 

As mentioned before, our simplified analytic model does not track the precise value of the homodyne current $x_c$, but only its sign. This corresponds to averaging over different outcomes with even (odd) qubit parity, leading to Eqn. \eqref{lind_eqn} making the quasi-stationary state $\rho_{e(o)}^{\rm{qs}}$ slightly impure. This leads to impurity in the post-measurement qubit state for a some regions in the outcome plane and is therefore an artifact of the approximation. The small regions of spurious zero concurrence in Fig. \ref{fig_x_y_meas} are due to this. As $\alpha, \beta$ are increased, the impurity due to the Lindblad evolution increases. This restricts the accuracy of the approximate analytical theory to $\alpha, \beta \sim 1$. For $\alpha, \beta \gg 1$, one can numerically simulate the evolution using the stochastic master equation. A comparison between the analytical solutions and stochastic master equation solutions is provided in \cite{Suppl} for $\alpha,\beta \sim {\cal O}(1)$. 

\begin{figure}
\includegraphics[width = 0.5\textwidth]{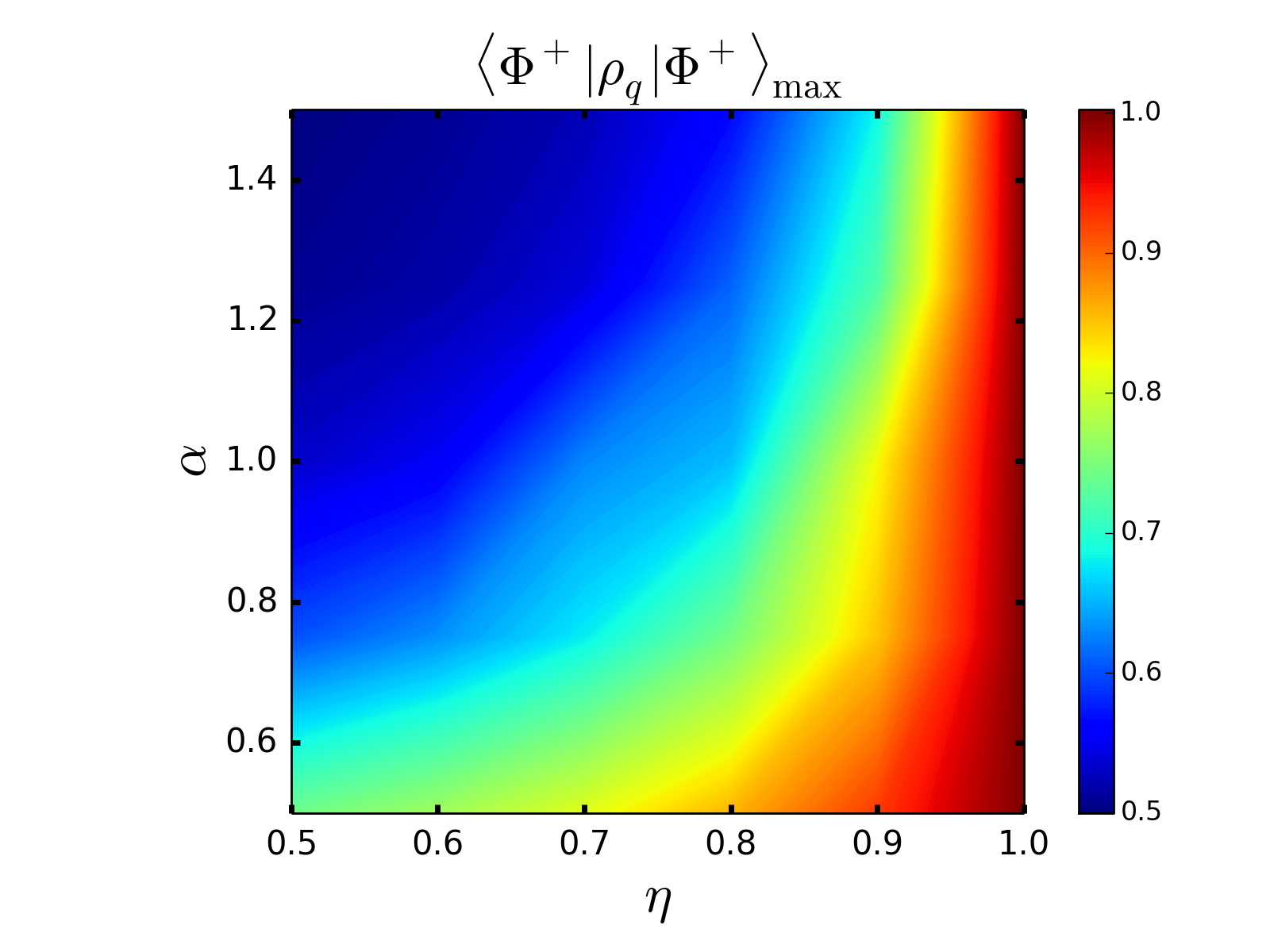}
\caption{ \label{fid_eta_alpha}  (color online) Maximum overlap with Bell state $|\Phi^+\rangle$ is shown upon variation of incident state amplitude $\alpha$ and efficiency parameter $\eta$ for measurement of Y quadratures of $\mathbf{a}$ and $\mathbf{b}$. A sample of $500$ trajectories were simulated for each data point in the $(\alpha,\eta)$ space and maximum fidelity was noted. For perfect efficiency ($\eta = 1$), it is always possible to generate  $|\Phi^+\rangle$. The maximum target fidelity goes down as $\eta$ is lowered. Higher values of $\alpha$ is more susceptible to finite efficiency. However, even for $\eta = 0.7$ and $\alpha = 0.5$, we obtain fidelities in excess of $80\%$, indicating the robustness of the scheme to photon loss and finite quantum efficiency.}
\end{figure}

In what follows, we test the robustness of our protocol to imperfections arising out of undesired photon loss and finite quantum efficiency. We treat these imperfections together as a general efficiency parameter $\eta$. 
We present here results of numerical simulations for the case when both Y quadratures were measured (Fig. \ref{fid_eta_alpha}). Similar results are obtained when X measurements are performed. While for perfect efficiency ($\eta = 1$), it is always possible to generate  an entangled Bell state, even for $\eta = 0.7$ and $\alpha = 0.5$, fidelity in excess of $80\%$ is obtained, indicating the robustness of the protocol to these imperfections. However, the photon losses before the JPM prevent a complete trade-off between the success-rate and fidelity as will be discussed in \cite{Roy}. Note that a low value of $\alpha, \beta$ lowers the success rate of entanglement generation since the two-qubit parity measurement in the step (II) relies on unambiguously inferring the location of the homodyne outcome of the parity signal. 

To summarize, we have presented a protocol for remotely entangling two qubits by performing a set of  concurrent quantum operations on propagating microwave modes entangled with the qubits. In contrast to existing schemes based on linear optical elements and photon detectors, we propose a qualitatively different approach, based on the multiplication of quantum signals prior to continuous measurements to generate remote entanglement. This multiplication is achieved using the Josephson nonlinearity and the high detection efficiency of microwave radiation in circuit QED systems promise a much higher success rate of entanglement generation compared to its optical counterparts. 

Discussions with Michael Hatridge, Zaki Leghtas, Matti Silveri and Steve Girvin are gratefully acknowledged. The work was supported by US Army Research Office Grant No. W911NF-14-1-0011 and NSF grant ECCS 1068642. LJ acknowledges the support of the DARPA Quiness program, Alfred P. Sloan Foundation and Packard Foundation.

\bibliography{references}

\end{document}